\documentclass[preprint,aps,12pt,showpacs,nofootinbib,tightenlines]{revtex4}
\usepackage{amsmath}
\usepackage{amssymb}
\usepackage{epsfig}
\usepackage{graphicx}
\begin{document}
%\preprint{NJNU-TH-07-05}
%%%%%%%%%%%%%%%%%%%%%%%%%%%%%%%%%%%%%%%%%%%%%
%%%%%%%%%%%%%%%%%%%%%%%%%%%%%%%%%%%%%%%%%%%%%%%%%%%%%%
\def\pslash{\rlap{\hspace{0.02cm}/}{p}}
\def\eslash{\rlap{\hspace{0.02cm}/}{e}}
%%%%%%%%%%%%%%%%%%%%%%%%%%%%%%%%%%%%%%%%%%%%%%%%%%%%%%
\title{Probing lepton flavor violation signal via
 $e^+e^-(\gamma\gamma)\rightarrow l_{i}\bar{l}_{j}$ in the
littlest Higgs model with T-parity at the ILC}

\author{Jinzhong Han}
\author{Xuelei Wang}\email{wangxuelei@sina.com}
\author{Bingfang Yang}
\affiliation{  College of Physics and Information Engineering, Henan
Normal University, Xinxiang 453007, China \vspace*{1.5cm}}

%\date{\today}
\begin{abstract}
In the littlest Higgs model with T-parity, the new interactions
between the mirror leptons and the Standard Model leptons can induce
some lepton flavor violation (LFV) processes at loop level. We study
the possibility of the ILC to probe the LFV production processes
$e^+e^-(\gamma\gamma)\rightarrow l_{i}\bar{l}_{j}$. Our results show
that the rates of $\gamma\gamma\rightarrow l_{i}\bar{l}_{j}$ can
reach 1 fb in optimal cases after reasonable kinematical cuts, which
implies that these processes may be observed at the ILC.
\end{abstract}


\maketitle

%%%%%%%%%%%%%%%%%%%%%%%%%%%%%%%%%%%%%%%%%%%%%%%%%%%%%%%%%%%%%%%%%%%%%%%%%%%%%%%

\section{ Introduction}
The little Higgs theory is proposed as an elegant solution to the
hierarchy problem of the Standard Model (SM) and is now an important
candidate of new physics \cite{little Higgs}. Among various little
Higgs models, the Littlest Higgs (LH) model \cite{LH} is the
simplest but phenomenologically viable model, which incorporates all
essential ingredients of the little Higgs theory. Unfortunately,
this economic model suffers from severe constraints from the
precisely measured electroweak data, and one has to tune finely its
parameters to survive the constraints \cite{constraints}. To avoid
this problem, a new discrete symmetry called T-parity was introduced
and the resulting model is referred to as the Littlest Higgs model
with T-parity (LHT) \cite{LHT}. In the LHT model, all dangerous
contributions to the electroweak data are loop suppressed, and
consequently wide regions of its parameter space are consistent with
the data even when the breaking scale of the T-parity, $f$,  is as
low as $500$ GeV \cite{EW constraint}.  On the other hand, to
implement the T-parity one has to introduce a mirror fermion (T-odd
quark/lepton) for each SM fermion.  In general, the mass matrix for
the mirror quarks/leptons is not proportional to that of their SM
counterparts, and due to the misalignment of the mass matrix,
neutral flavor changing (FC) interactions between the two types of
fermions may naturally appear, which will induce flavor changing
neutral current (FCNC) processes for both quarks and leptons at loop
level \cite{FC-LHT3,feynman rule,FC-LHT0,quark-LHT}. Since these
processes are highly suppressed in the SM, they can be utilized to
probe new physics and any observation of them will undoubtedly imply
the existence of new physics.

Since the observation of the neutrino oscillation, searching for the
LFV signals at colliders has attracted more and more attention
\cite{experiment1,experiment2,LFV-SUSY,LFV-TC2}. The LFV production
processes, such as $p p(\bar{p})\rightarrow l_{i}\bar{l}_{j}$
\cite{RPV pp} and $\gamma\gamma\rightarrow l_{i}\bar{l}_{j}$,  have
been studied in R-parity conversing MSSM \cite{RPV pp,RPC MSSM},
R-parity violating MSSM \cite{RPV MSSM} and TC2 model \cite{rr-ll
TC2}. These study indicates that the production rates can be several
order larger than those in the SM and may reach the sensitivity of
future experiments. We note that, in the LHT the FC interaction can
also induce the production processes, and compared  the LFV decays
such as $l_{i}\rightarrow l_{j}\gamma$, $l_{i}\rightarrow
l_{j}l_{k}l_{l}$, $\tau\rightarrow \mu\pi$ studied in the LHT model
\cite{g-2u,LHT-lilj,LFV-LHT} with the decays in other new physics
models \cite{RPC MSSM,RPV MSSM}, we infer that the production rate
is not suppressed in comparison with the predictions of the other
model. This encourages us to study the production processes in
detail. Among different production processes, we are more interested
in $e^+e^-(\gamma\gamma)\rightarrow l_{i}\bar{l}_{j}$ ($i\neq j$ and
$l_i=e,\mu,\tau$) occurred at the proposed International Linear
Collider (ILC) since the ILC provides rather clean environment to
probe these processes.

\indent  This paper is organized as follows. In Section II we
briefly review the LHT model. In Section III and IV, we show the
details of our calculation of the production rates and present
numerical results respectively. Finally, a short conclusion is drawn
in Section V.

\section{ Review of the LHT model}

In this sector, we briefly recapitulate the structure of the LHT
model and define the conventions of our notation.  A detailed
description of the model can be found in \cite{FC-LHT3,feynman
rule}.

Basically speaking, the LHT model is a non-linear sigma model
describing the spontaneous breaking of a global $SU(5)$ down to a
global $SO(5)$. This symmetry breaking takes place at the scale
$f\sim\mathcal {O}$ (TeV), and along with this breaking, there arise
14 Goldstone bosons which are described by the ``pion"matrix
\begin {equation}
\Pi=
\begin{pmatrix}
-\frac{\omega^0}{2}-\frac{\eta}{\sqrt{20}}&-\frac{\omega^+}{\sqrt{2}}
&-i\frac{\pi^+}{\sqrt{2}}&-i\phi^{++}&-i\frac{\phi^+}{\sqrt{2}}\\
-\frac{\omega^-}{\sqrt{2}}&\frac{\omega^0}{2}-\frac{\eta}{\sqrt{20}}
&\frac{v+h+i\pi^0}{2}&-i\frac{\phi^+}{\sqrt{2}}&\frac{-i\phi^0+\phi^P}{\sqrt{2}}\\
i\frac{\pi^-}{\sqrt{2}}&\frac{v+h-i\pi^0}{2}&\sqrt{4/5}\eta&-i\frac{\pi^+}{\sqrt{2}}&
\frac{v+h+i\pi^0}{2}\\
i\phi^{--}&i\frac{\phi^-}{\sqrt{2}}&i\frac{\pi^-}{\sqrt{2}}&
-\frac{\omega^0}{2}-\frac{\eta}{\sqrt{20}}&-\frac{\omega^-}{\sqrt{2}}\\
i\frac{\phi^-}{\sqrt{2}}&\frac{i\phi^0+\phi^P}{\sqrt{2}}&\frac{v+h-i\pi^0}{2}&-\frac{\omega^+}{\sqrt{2}}&
\frac{\omega^0}{2}-\frac{\eta}{\sqrt{20}}   \label{Goldstone}
\end{pmatrix}.
\end{equation}
Among the Goldstone bosons,  the fields $\omega^0, \omega^\pm$ and
$\eta$ are eaten by the new T-odd heavy gauge bosons $Z_H$,
$W_H^{\pm}$ and $A_H$. As a result, these gauge bosons acquire
masses which up to $\mathcal {O}(v^2/f^2)$ are given by
\begin{equation}
M_{W_{H}^{\pm
}}=M_{Z_{H}}=gf(1-\frac{v^2}{8f^2}),~~~~~~~~M_{A_H}=\frac{g'}{\sqrt{5}}f(1-\frac{5v^2}{8f^2})
\label{heavy-gauge-boson}
\end{equation}
with $g$ and $g'$ denoting the SM $SU(2)$ and $U(1)$ gauge couplings
respectively and $v$ being the electroweak breaking scale. Quite
similarly, the fields $\pi^0 $ and $\pi^\pm$ are eaten by the SM
gauge bosons (T-even) after the electroweak symmetry breaking. The
masses of the $Z$ and $W$ bosons are then given by
\begin{equation}
M_{W_{L}}=\frac{gv}{2}(1-\frac{v^2}{12f^2}),~~~~~~~~M_{Z_L}=\frac{gv}{2\cos\theta_W}(1-\frac{5v^2}{12f^2}).
\label{SM-gauge-boson}
\end{equation}
Note in the LHT model, the neutral gauge boson $A_H$ is the lightest
T-odd particle, and due to the conservation of T-parity, it is
stable and thus can act as an dark matter candidate \cite{EW
constraint}.

In order to implement the T-parity, each SM fermion must be
accompanied by its mirror fermion. The particle content of the LHT
then includes the T-even fermions, such as the SM quarks, leptons
and an additional heavy quark $T_{+}$, and their mirror fermions. In
this paper, we denote the mirror leptons by
\begin{eqnarray}
\left(
  \begin{array}{c}
    l_{H}^{1} \\
    \nu_{H}^{1} \\
  \end{array}
\right),~~~
 \left(
  \begin{array}{c}
    l_{H}^{2} \\
    \nu_{H}^{2} \\
  \end{array}
\right),~~~
 \left(
  \begin{array}{c}
    l_{H}^{3} \\
   \nu_{H}^{3} \\
  \end{array}
\right).
\end{eqnarray}
with their masses given by \cite{FC-LHT3,feynman rule}
\begin{eqnarray} m_{l_H^i}=\sqrt{2}\kappa_if\equiv m_{H_i},
~~~m_{\nu_H^i}= m_{H_i}(1-\frac{v^2}{8f^2})
\end{eqnarray}
Obviously, neglecting the $\mathcal {O}(v^2/f^2)$ correction to
$m_{\nu_H^i}$ ($i$ is generation index), the mirror neutrino and the
mirror lepton in the same generation are degenerated in mass.

In a similar way to what happens for the SM fermions, the mirror
sector has weak mixing parameterized by unitary mixing matrices,
i.e. $V_{H_l}$, $V_{H_{\nu}}$ for mirror leptons and $V_{H_u}$,
$V_{H_d}$ for mirror quarks which satisfy the following physical
constraints:
\begin{equation}
V^{\dag}_{H_\nu}V_{H_l}=V_{PMNS}, ~~V^{\dag}_{H_u}V_{H_d}=V_{CKM}.
\end{equation}
These mirror mixing matrices  imply flavor violating interactions
between SM fermions and mirror fermions that are mediated by the
heavy gauge boson $W_H$, $Z_H$ and $A_H$. The relevant Feynman rules
are given in \cite{FC-LHT0}.
\begin{figure}
\begin{center}

\includegraphics [scale=0.50] {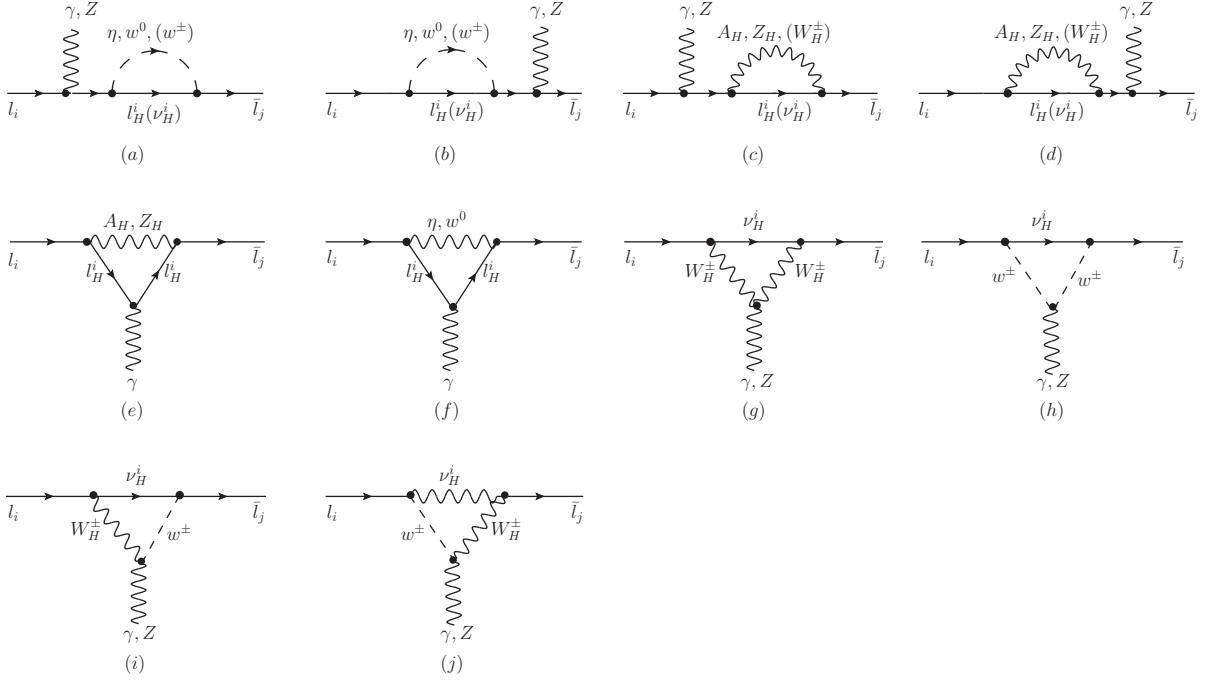}
\caption{Feynman diagrams contributing to the LFV vertex
$l_i\bar{l}_jZ(\gamma)$ in the LHT model.} \label{fig:fig1}
\end{center}
\end{figure}

\section{Calculations}
\subsection{The LFV interaction $ \bar{l}_i l_jV(V=\gamma,Z)$
in the LHT model}

We have mentioned that the interaction between the mirror lepton and
the SM lepton, such as $\bar{l}_H l Z_H (A_H) $ and $\bar{\nu}_H l
W_H$, can induce LFV interactions at loop level.  The relevant
Feynman diagrams are shown in Fig.\ \ref{fig:fig1} for $\bar{l}_i
l_j Z$ and $\bar{l}_i l_j \gamma$ vertexes. Unlike the previous
studies where the unitary gauge were used in the calculation
\cite{g-2u,LHT-lilj}, we use Feynman gauge to obtain our results,
and that is why we also plot the diagrams involving the Goldstone
bosons $\eta$, $\omega^0$ and $\omega^{\pm}$ in Fig.\
\ref{fig:fig1}. We once compared our results for the decay $l_i \to
l_j \gamma$ with \cite{LHT-lilj} and found we can reproduce Fig.\ 4
of this literature. Note in the LHT, the T-odd scalar triplet $\Phi$
in Eq.\ (\ref{Goldstone}) can also contribute to the LFV vertex by
the $\bar{l}_H l \Phi$ interaction. However, since such interaction
is suppressed by ${v^2}/{f^2}$, we can neglect its contribution at
the leading order of ${v}/{f}$ expansion.

\begin{figure}
\begin{center}
\includegraphics [scale=0.5] {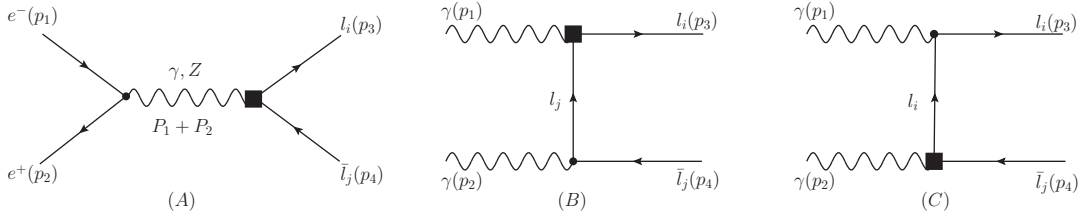}
\caption{Feynman diagrams for the production
$e^+e^-(\gamma\gamma)\rightarrow l_i\bar{l}_j$ in the LHT model with
the black squares denoting the effective $\bar{l}_i l_j Z(\gamma)$
vertex introduced in \cite{effective vertex method}. Diagrams with
the two photon lines crossed are not shown for $\gamma\gamma
\rightarrow l_i\bar{l}_j$. } \label{fig:fig2}
\end{center}
\end{figure}

The Feynman diagrams for the production $e^+ e^- (\gamma \gamma)
\rightarrow l_{i}\bar{l}_{j}$ are shown in Fig.\ \ref{fig:fig2} with
the black square denoting the loop-induced $\bar{l}_i l_j Z(\gamma)
$ vertex. One important difference of the $\bar{l}_i l_j \gamma $
vertex in $e^+ e^- \rightarrow l_{i}\bar{l}_{j}$ and in $\gamma
\gamma \rightarrow l_{i}\bar{l}_{j}$ is both the leptons are
on-shell for $e^+ e^- \rightarrow l_{i}\bar{l}_{j}$, while either
$l_i$ or $l_j$ is off-shell  for $\gamma \gamma \rightarrow
l_{i}\bar{l}_{j}$. In order to simplify calculation, we'd better use
an universal form of the $\bar{l}_i l_j \gamma$ vertex which is
valid for the both cases. This is possible as suggested by
\cite{effective vertex method}. In our calculation, we use the
method in \cite{effective vertex method} to get the effective
$\bar{l}_i l_j Z(\gamma)$ vertex and present their expressions in
detail in Appendix A. We numerically check the rates of the
production processes are free of ultraviolet divergence. We use the
code LoopTools \cite{LoopTools} to calculate the loop functions
appeared in the effective vertexes.

\subsection{Amplitudes for $e^+e^-(\gamma\gamma)\rightarrow l_{i}\bar{l}_{j}$}
With the aid of the effective $\bar{l}_i l_j Z(\gamma)$ vertex, one
can write down the amplitude of $e^+e^-\rightarrow l_{i}\bar{l}_{j}$
by a straightforward calculation of Fig.\ \ref{fig:fig2}(A):
\begin{eqnarray}
M_A=M^{\gamma}_A+M^{Z}_A,
\end{eqnarray}
with
\begin{eqnarray}
 M_A^{\gamma}&=& - \frac{e}{(p_1+p_2)^2} \bar{u}_{l_i}(p_{3})\Gamma^{\mu}_{\bar{l}_i l_j \gamma} (p_{3},p_{4})v_{\bar{{l}}_{j}}(p_{4})\bar{v}_{e^{+}}(p_{2})\gamma_{\mu}u_{e^{-}}
 (p_{1}), \nonumber \\
 M_A^{Z}&=&\frac{g}{\cos\theta_W}\frac{1}{(p_1+p_2)^2-M_Z^2}\bar{u}_{l_i}(p_{3})\Gamma^{\mu}_{\bar{l}_i l_j Z}(p_{3},p_{4}) v_{\bar{{l}}_{j}}(p_{4})\bar{v}_{e^{+}}(p_{2})\gamma_{\mu}\\
 \nonumber
 &&\times[(-\frac{1}{2}+\sin^{2}\theta_W)P_{L}+(\sin^{2}\theta_W)P_{R}]u_{e^{-}}(p_{1}),
\end{eqnarray}
where $\Gamma^{\mu}_{\bar{l}_i l_j Z}$ ($\Gamma^{\mu}_{\bar{l}_i l_j
\gamma}$) is the effective $\bar{l}_i l_j Z$($\bar{l}_i l_j \gamma
$) vertex which depends on the lepton momenta $p_3$ and $p_4$,
$P_L=\frac{1}{2}(1-\gamma_5)$ and $P_R=\frac{1}{2}(1+\gamma_5)$.

Similarly, the amplitude of $\gamma \gamma \rightarrow
l_{i}\bar{l}_{j}$ is given by
\begin{eqnarray}
M_B&=&\frac{e}{(p_3-p_1)^2 - m_{l_j}^2} \bar{u}_{l_i}(p_{3})
\Gamma^{\mu}_{\bar{l}_i l_j \gamma} (p_3,p_1-p_3)
\epsilon_{\mu}(p_1)(\pslash_3-\pslash_1+m_{\bar{l}_j})
\rlap/\epsilon(p_2) v_{{\bar{l}_j}}(p_{4}),  \nonumber \\
M_C&=&\frac{e}{(p_2-p_4)^2
-m_{l_i}^2}\bar{u}_{l_{i}}(p_{3})\rlap/\epsilon(p_1)(\pslash_2-\pslash_4+m_{l_{i}})
\Gamma^{\mu}_{\bar{l}_i l_j
\gamma}(p_{2}-p_{4},p_4)\epsilon_{\mu}(p_{2})
v_{{\bar{l}_j}}(p_{4}).
 \end{eqnarray}

For the $\gamma \gamma$ collision at the ILC, the photon beams are
generated by the backward Compton scattering of incident electron-
and laser-beams just before the interaction point. The events number
is obtained by convoluting the cross section with the photon beam
luminosity distribution. For $\gamma \gamma$ collider the events
number is obtained by
\begin{eqnarray}
N_{\gamma \gamma \to  l_{i}\bar{l}_{j}}&=&\int  {\rm d}
\sqrt{s_{\gamma \gamma}} \frac{{\rm d}{\cal L}_{\gamma \gamma}}
{{\rm d} \sqrt{s_{\gamma \gamma}}} \hat{\sigma}_{\gamma \gamma \to
l_{i}\bar{l}_{j}} (s_{\gamma \gamma})\equiv {\cal L}_{e^+e^-}
~\sigma_{\gamma \gamma \to  l_{i}\bar{l}_{j}}(s_{e^+e^-}),
\label{definition}
\end{eqnarray}
where ${\rm d}{\cal L}_{\gamma \gamma}/{\rm d} \sqrt{s_{\gamma
\gamma}}$ is the photon beam luminosity distribution and
$\sigma_{\gamma \gamma \to  l_{i}\bar{l}_{j}}(s_{e^+e^-}) $, with
$s_{e^+e^-}$ being the energy-square of $e^+e^-$ collision, is
defined as the effective cross section of $ \gamma \gamma \to
l_{i}\bar{l}_{j}$.  In optimum case, $\sigma_{\gamma \gamma \to
l_{i}\bar{l}_{j}} $ can be written as \cite{distribution}
\begin{eqnarray}
\sigma_{\gamma \gamma \to
l_{i}\bar{l}_{j}}(s_{e^+e^-})&=&\int_{\sqrt{a}}^{x_{max}} 2 z{\rm d}
z
 ~\hat{\sigma}_{\gamma \gamma \to  l_{i}\bar{l}_{j}} (s_{\gamma \gamma}=z^2 s_{e^+e^-})
\int_{z^2/x_{max}}^{x_{max}} \frac{{\rm d} x}{x}~F_{\gamma/e}(x)
~F_{\gamma/e}(\frac{z^2}{x}), \label{cross}
\end{eqnarray}
where $F_{\gamma/e}$ denotes the energy spectrum of the
back-scattered photon for unpolarized initial electron and laser
photon beams given by
\begin{eqnarray}
F_{\gamma/e}(x)&=&\frac{1}{D(\xi)} \left ( 1-x+\frac{1}{1-x}-\frac{4
x}{\xi (1-x)}+ \frac{4 x^2}{\xi^2 (1-x)^2} \right ) .
\end{eqnarray}
The definitions of parameters $\xi$, $D(\xi)$ and $x_{max}$ can be
found in \cite{distribution}. In our numerical calculation, we
choose $\xi=4.8$, $D(\xi)=1.83$ and $x_{max}=0.83$.

Before we end this section, we emphasize two advantages of $\gamma
\gamma$ collision over the $e^+ e^-$ collision of the ILC in probing
the LFV interaction. One is for the process $e^+ e^- \rightarrow
l_{i}\bar{l}_{j}$, it occurs only via s-channel, so its rate is
suppressed by the photon propagator and the $Z$ propagator. While
for $\gamma \gamma \rightarrow l_{i}\bar{l}_{j}$, there is no such
suppression. The other is the $\gamma \gamma$ collision provides a
cleaner environment than the $e^+e^-$ collision, so is well suited
to probe new physics.

\section{Numerical results}
In our calculations, we neglect terms proportional to $v^2/f^2$ in
the new gauge boson masses and also in the relevant Feynman rules.
We take the SM parameters as \cite{parameters}:
\begin{eqnarray}
&& m_{e}=0.0051\ {\rm GeV}, \quad \quad  m_{\mu}= 0.106\ {\rm GeV},
\quad \quad  m_\tau= 1.777\ {\rm GeV}, \nonumber \\ &&  m_Z =91.2\
{\rm GeV}, \quad \quad \quad s^{2}_{W}= 0.231, \quad \quad
\alpha_e=1/128.8. \nonumber
\end{eqnarray}
Among the LHT parameters, we must specify the breaking scale $f$,
the mirror lepton masses as well as  the matrix $ V_{H_l}$. We
choose $f=500 $ GeV and $f=1000 $ GeV as two representative cases,
and as shown in \cite{EW constraint}, these two cases are consistent
with precision electroweak data. About the mirror lepton masses, we
fix $m_{l_H^1}=m_{l_H^2}=m_{H12}=500 $ GeV, and vary
$m_{l_H^3}\equiv m_{H3}$ in the range of $600-1200$ GeV for $f=500$
GeV and $600-1500$ GeV for $f=1000$ GeV. We set $V_{Hl} = V_{PMNS}$
(or equally $V_{H\nu}=I$),  and like \cite {feynman rule,g-2u} did,
determine the elements of $V_{PMNS}$ from the neutrino experiments
\cite{Vpmns-1,Vpmns-2,Vpmns-3,Vpmns-4,Vpmns-5} with the three
Majorana phases in $V_{PMNS}$ taken to be zero. From the results of
\cite{g-2u}, one can learn that our choice of the LHT parameters
satisfies the constraint from $m_{H_i}\leq 4.8f^2$ and the rare
decays $l_{i}\rightarrow {l_{j}}\gamma$. By the way, in calculating
the rates for $\gamma \gamma \to l_{i}\bar{l}_{j}$ we require
$|\cos\theta_l|<0.9$ and $p^l_T
> 20$ GeV \cite{RPC MSSM}.

\begin{figure}[th]
\scalebox{0.7}{\epsfig{file=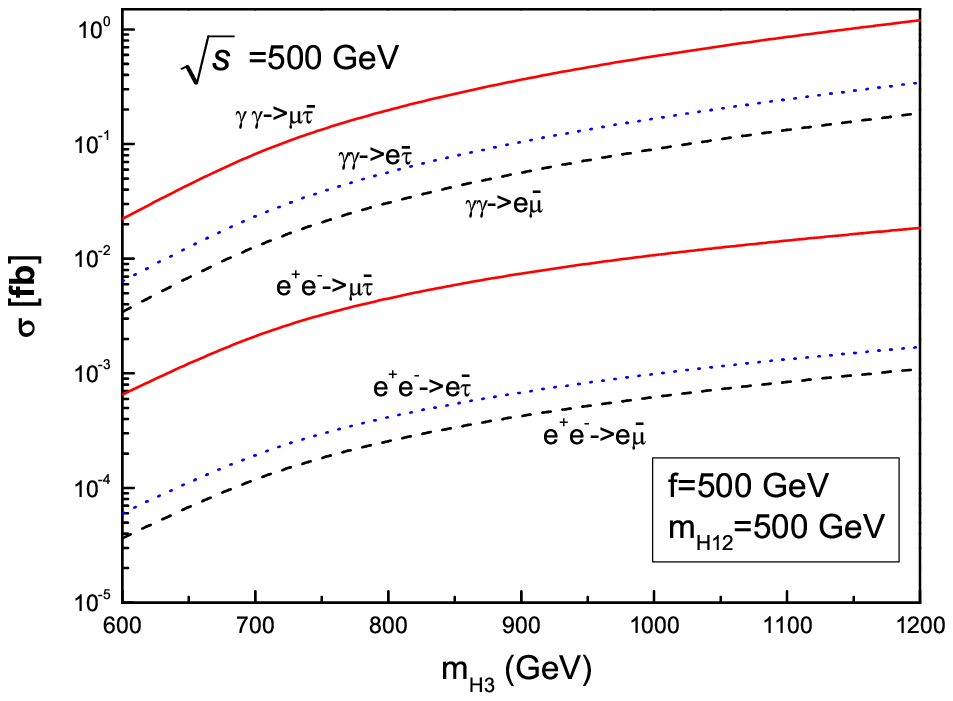}}
\scalebox{0.7}{\epsfig{file=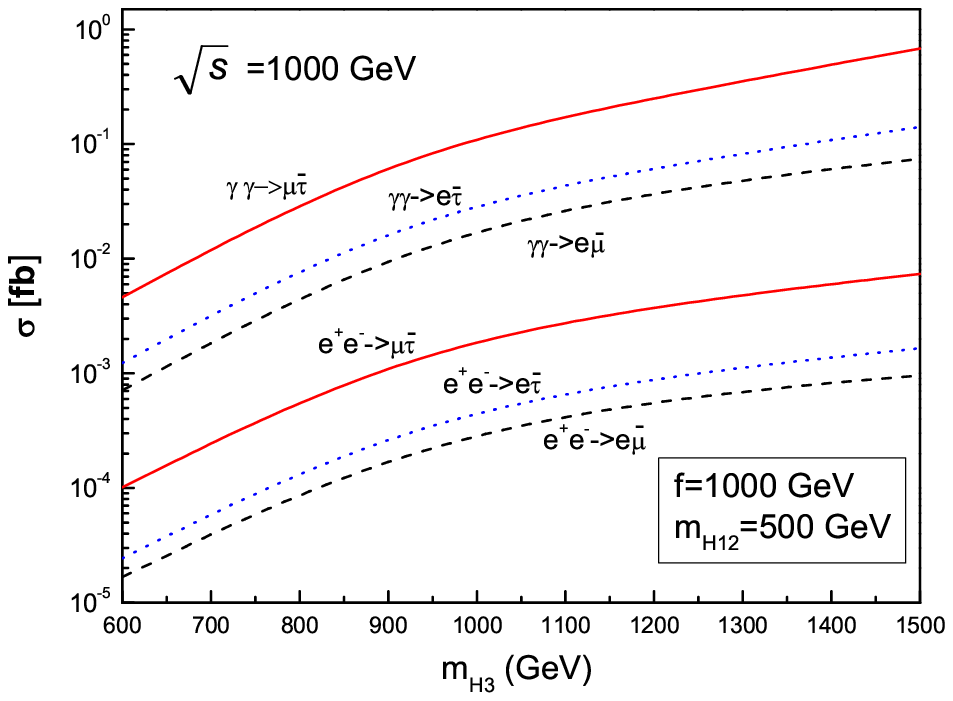}}\\
\caption{\small The production rates for the processes
$e^+e^-(\gamma\gamma)\rightarrow l_{i}\bar{l}_{j}$ as a function of
$m_{H3}$.}
\end{figure}

Our results are summarized in Fig. 3 with different $l_i \bar{l}_j$
states considered. From this figure, one can get three conclusions.
The first is the production rates of
$e^+e^-(\gamma\gamma)\rightarrow l_{i}\bar{l}_{j}$ monotonously
increase with $m_{H3}$ becoming larger. This is because these
processes proceed in a way quite similar to the GIM mechanism of the
SM, so the more significant the mass splitting between the mirror
leptons is, the larger the rates become. The second is the rate for
$\gamma\gamma\rightarrow l_{i}\bar{l}_{j}$ is several orders larger
than that of $e^+e^-\rightarrow l_{i}\bar{l}_{j}$. The reason is, as
we mentioned before, that the process $e^+e^-\rightarrow
l_{i}\bar{l}_{j}$ is s-channel suppressed, while the process
$\gamma\gamma\rightarrow l_{i}\bar{l}_{j}$ gets contribution from
u-channel and t-channel and there is no such suppression. The last
is among the LFV processes, the rate for $e\mu$ final state is much
smaller than that for $e \tau$ or $\mu \tau$ state. This is because
the processes with $e\mu$ final state are stringently constrained by
the decay $\mu\rightarrow e\gamma$.

\begin{figure}
\scalebox{0.7}{\epsfig{file=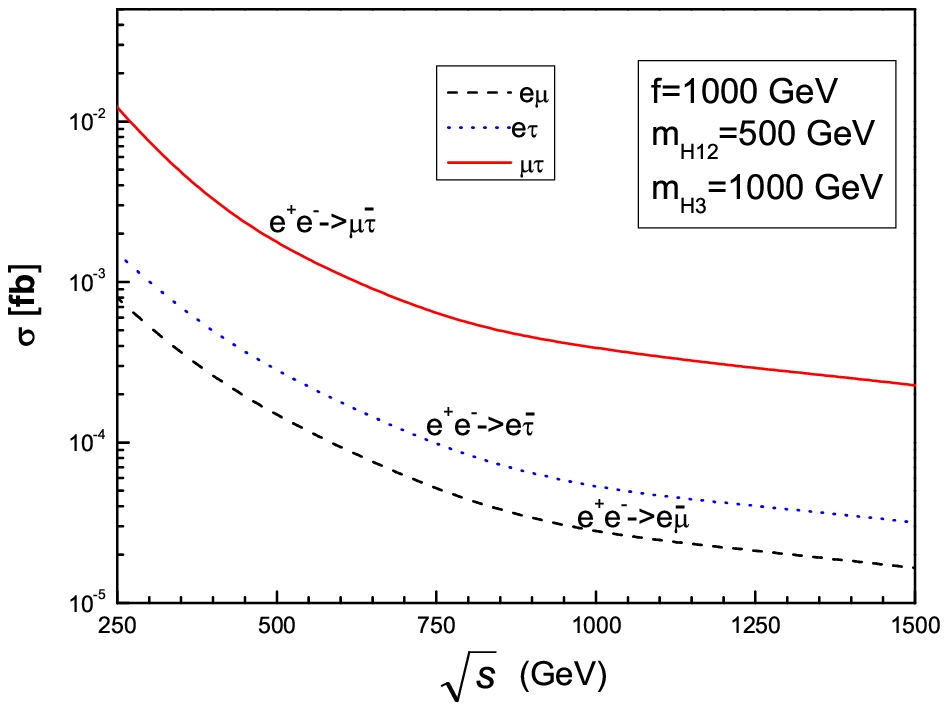}}
\scalebox{0.7}{\epsfig{file=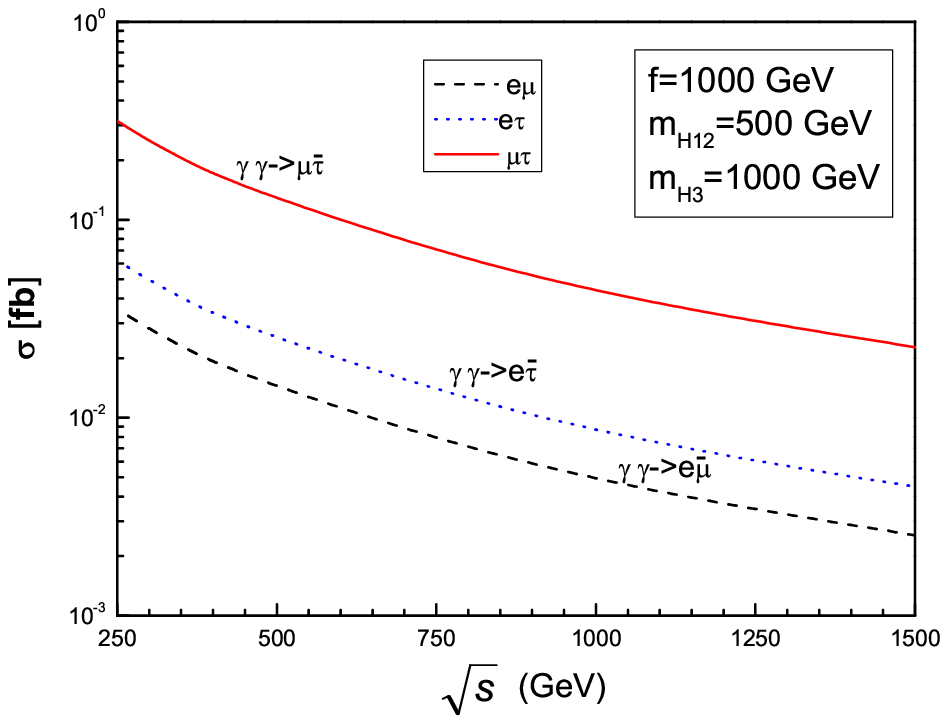}}\\
\caption{\small The production rates for the processes
$e^+e^-(\gamma\gamma)\rightarrow l_{i}\bar{l}_{j}$ as a function of
center-of-mass energy $\sqrt{s}$.}
\end{figure}

Since the production rates for $\gamma\gamma\rightarrow e\tau,
\mu\tau$ are significantly larger than the other production rates,
we now discuss their observability at the ILC. For
$\gamma\gamma\rightarrow e\tau $, its main backgrounds come from
$\gamma\gamma\rightarrow \tau^+\tau^-\rightarrow
\tau^-\nu_e\bar{\nu}_\tau{e^+}$, $\gamma\gamma\rightarrow
W^+W^-\rightarrow \tau^-\nu_e\bar{\nu}_\tau{e^+}$ and
$\gamma\gamma\rightarrow e^+e^-\tau^+\tau^-$. In order to enhance
the ratio of the signal to the background, one usually adds the
following cuts in Monte Carlo simulation \cite{RPC MSSM}:
$|\cos\theta_l|<0.9$ and $p^e_T
> 20$GeV.  With these cuts, the rates for the background
processes at $\sqrt{s}$= 500GeV are $9.7\times10^{-4}$ fb for
$\gamma\gamma\rightarrow \tau^+\tau^-\rightarrow
\tau^-\nu_e\bar{\nu}_\tau{e^+} $, $1.0\times10^{-1}$ fb for
$\gamma\gamma\rightarrow W^+W^-\rightarrow
\tau^-\nu_e\bar{\nu}_\tau{e^+}$, and $2.4\times10^{-2}$ fb for
$\gamma\gamma\rightarrow e^+e^-\tau^+\tau^-$ respectively (see Table
1 of \cite{RPC MSSM}). This implies that to get a $3\sigma$
observing sensitivity with $3.45\times{10^{-2}}$ fb$^{-1}$
integrated luminosity \cite{luminosity}, the production rate for
$\gamma\gamma\rightarrow e\bar{\tau}$ must be larger than
$2.5\times10^{-2}$ fb  \cite{RPC MSSM}.  Compared this value with
the results in Fig.\ 3, one can learn that the process $\gamma
\gamma \to e \bar{\tau}$ may be observable in broad regions of the
LHT parameter space. With regard to $\gamma \gamma \to \mu
\bar{\tau}$, since its production rate may be several times larger
than that for $\gamma \gamma \to e \bar{\tau}$ while the background
rates are same, one can conclude that $\gamma \gamma \to \mu
\bar{\tau}$ is more powerful in probing the LHT model.

For the sake of providing more information of the ILC in probing the
LHT model, we also show the rates of
$e^+e^-(\gamma\gamma)\rightarrow l_{i}\bar{l}_{j}$ as the function
of center-of-mass energy of the ILC $\sqrt{s}$ in Fig.\ 4. We see
that with the increase of $\sqrt{s}$, the production rates become
smaller which is similar to the behaviors of the supersymmetric
models \cite{RPC MSSM,RPV MSSM}.

We also list the theoretical predictions of the production rates in
the optimum case of different models in Table I. From this table,
one can learn that, due to enhanced coupling strength, the
production rates in the LHT and TC2 model can be significantly
larger than that in R-parity violating MSSM. So the processes of
$\gamma\gamma\rightarrow l_{i}\bar{l}_{j}$ may be utilized to
distinguish new physics models.
\begin{table}

\caption{\small The theoretical predictions of the rates for $\gamma
\gamma \to l_{i}\bar{l}_{j}$ at $\sqrt{s}=500{\rm }$ GeV in the
optimum case of different models.}

\vspace{0.3cm}

\begin{tabular}{|c|c|c|c|c|c|c|}\hline
&MSSM with R-parity &MSSM without R-parity &TC2&LHT\\
 \hline
  $\gamma\gamma
\rightarrow \mu\bar{\tau}$     &$\mathcal {O}(10^{-2})$\cite{RPC MSSM} &$\mathcal {O}(10^{-2})$\cite{RPV MSSM}&$\mathcal {O}(1)$\cite{rr-ll TC2}&$O(1)$  \\
\hline
  $\gamma\gamma
\rightarrow e\bar{\tau}$    &$\mathcal {O}(10^{-1})$\cite{RPC MSSM} &$\mathcal {O}(10^{-2})$\cite{RPV MSSM}&$\mathcal {O}(1)$\cite{rr-ll TC2}&$O(10^{-1})$  \\
\hline $\gamma\gamma
\rightarrow e\bar{\mu}$  &$\mathcal {O}(10^{-3})$ \cite{RPC MSSM}&$\mathcal {O}(10^{-4})$\cite{RPV MSSM}&$\mathcal {O}(10^{-3})$\cite{rr-ll TC2}&$O(10^{-1})$  \\
 \hline
\end{tabular}
\end{table}

\section{Conclusion}

In the LHT model, the interactions between the mirror leptons and
the SM leptons induce the LFV processes at loop level. We study the
LFV productions $e^+e^-(\gamma\gamma)\rightarrow l_{i}\bar{l}_{j}$
at the ILC, and find that, compared with the SM predictions, the
production rates in the LHT can be greatly enhanced. In particular,
the production rates for $\gamma \gamma \to \mu \bar{\tau} $ and for
$\gamma \gamma \to e \bar{\tau} $ can reach 1 fb  and $10^{-1} $ fb
respectively in optimum case, which fall within the 3$\sigma$
observing sensitivity of the ILC. Therefore, these LFV production
process at the ILC may be utilized to probe the LHT model.

\section{Acknowledgments}
\hspace{1mm}

We would like to thank Junjie Cao  for helpful discussions and
kindly improving our manuscript. This work is supported by the
National Natural Science Foundation of China under Grant
Nos.10775039, 11075045, by Specialized Research Fund for the
Doctoral Program of Higher Education under Grant No.20094104110001
and by HASTIT under Grant No.2009HASTIT004.

\newpage
{\Large Appendix A: Explicit expressions of
$\Gamma^{\mu}_{\bar{e}\mu\gamma} $ and $\Gamma^{\mu}_{\bar{e}\mu
Z}$}

\vspace{1cm}

In this appendix, we list the explicit expressions for the effective
$\bar{e}\mu \gamma$ ($\bar{e}\mu Z$) vertex
$\Gamma^{\mu}_{\bar{e}\mu \gamma} $($\Gamma^{\mu}_{\bar{e}\mu Z} $).
These expressions are obtained by a straightforward calculation of
Fig.\ \ref{fig:fig1}. In our calculation, we neglect terms
proportional to $v^2/f^2$, which appear in the new gauge boson
masses and also in the relevant Feynman rules. Other effective
vertices such as $\bar{e}\tau(\bar{\mu}\tau)\gamma $ and
$\bar{e}\tau(\bar{\mu}\tau)Z $ can be obtained in a similar way.
\begin{eqnarray*}
&&\Gamma^{\mu}_{\bar{e}\mu\gamma}(p_e,p_{\bar{\mu}})
=\Gamma^{\mu}_{\bar{e}\mu\gamma}(\eta)+\Gamma^{\mu}_{\bar{e}\mu\gamma}(\omega^{0})
+\Gamma^{\mu}_{\bar{e}\mu\gamma}(\omega^{\pm})+\Gamma^{\mu}_{\bar{e}\mu\gamma}(A_{H})+\Gamma^{\mu}_{\bar{e}\mu\gamma}(Z_{H})
+\Gamma^{\mu}_{\bar{e}\mu\gamma}(W_{H}^{\pm})\\
&&\quad \qquad\ \qquad\ \quad
+\Gamma^{\mu}_{\bar{e}\mu\gamma}(W_{H}^{\pm}\omega^{\pm}),
\nonumber \\
&&\Gamma^{\mu}_{\bar{e}\mu\gamma}(\eta)=-\frac{i}{16\pi^{2}}\frac{eg^{\prime2}}
{100M_{A_{H}}^{2}}
(V_{Hl})_{ie}^*(V_{Hl})_{i\mu}(A+B+C) \\
&&  \qquad\  A=\{ \frac{1}{p_{e}^{2}-m_{\mu}^{2}}[
m_{Hi}^{2}(m_{\mu}^{2}B_{0}^{a}+p_{e}^{2}B_{1}^{a})\gamma^{\mu}P_{L}
+m_{e}m_{\mu}(m_{Hi}^{2}B_{0}^{a}+p_{e}^{2}B_{1}^{a})\gamma^{\mu}P_{R}\\
&&\quad \quad \quad
\quad+m_{e}(m_{Hi}^{2}B_{0}^{a}+m_{\mu}^{2}B_{1}^{a})\pslash_e\gamma^{\mu}P_{L}
+m_{Hi}^{2}m_{\mu}(B_{0}^{a}+B_{1}^{a})\pslash_e\gamma^{\mu}P_{R}]\}\\
&&\quad \quad \ B=\{\frac{1}{p_{\bar{\mu}}^{2}-m_{e}^{2}}[m_{Hi}
^{2}(m_{e}^{2}B_{0}^{b}+p_{\bar{\mu}}^{2}B_{1}^{b})\gamma^{\mu}P_{L}
+m_{e}m_{\mu}(m_{Hi}^{2}B_{0}^{b}+p_{\bar{\mu}}^{2}B_{1}^{b})\gamma^{\mu}P_{R}\\
&&\quad \quad \quad\quad
-m_{e}m_{Hi}^{2}(B_{0}^{b}+B_{1}^{b})\gamma^{\mu}\pslash_{\bar{\mu}}P_{L}-m_{\mu}(m_{Hi}^{2}B_{0}^{b}
+m_{e}^{2}B_{1}^{b})\pslash_{\bar{\mu}}\gamma^{\mu}P_{R}]\} \\
&&\quad \quad \ C=\{[-m_{Hi}^{4}C_{0}^{1}\gamma^{\mu}P_{L}
-m_{e}m_{\mu}m_{Hi}^{2}C_{0}^{1}\gamma^{\mu}P_{R}
+m_{Hi}^{2}m_{e}C_{\alpha}^{1}\gamma^{\alpha}\gamma^{\mu}P_{L}
\\&&\quad\ \quad \quad \quad +m_{e}m_{Hi}^{2}(-\gamma^{\mu}\pslash_eC_{0}^{1}
-\gamma^{\mu}\pslash_{\bar{\mu}}C_{0}^{1}+\gamma^{\mu}\gamma^{\alpha}C_{\alpha}^{1})P_{L}
+m_{Hi}^{2}m_{\mu}C_{\alpha}^{1}\gamma^{\alpha}\gamma^{\mu}P_{R} \\
&&\quad\quad\ \quad
\quad+m_{Hi}^{2}m_{\mu}(\gamma^{\mu}\gamma^{\alpha}C_{\alpha}^{1}
-\gamma^{\mu}\pslash_eC_{0}^{1}-\gamma^{\mu}\pslash_{\bar{\mu}}C_{0}^{1})P_{R}
+m_{Hi}^{2}(\gamma^{\alpha}\gamma^{\mu}{\pslash_e}C_{\alpha}^{1}
+\gamma^{\alpha}\gamma^{\mu}{\pslash_{\bar{\mu}}}C_{\alpha}^{1} \\
&&\quad\quad\quad\
\quad-\gamma^{\alpha}\gamma^{\mu}\gamma^{\beta}C_{\alpha\beta}^{1})P_{L}
+m_{e}m_{\mu}(\gamma^{\alpha}\gamma^{\mu}{\pslash_e}C_{\alpha}^{1}
+\gamma^{\alpha}\gamma^{\mu}{\pslash_{\bar{\mu}}}C_{\alpha}^{1}-
\gamma^{\alpha}\gamma^{\mu}\gamma^{\beta}C_{\alpha\beta}^{1})P_{R}]
\},\\&&
\Gamma^{\mu}_{\bar{e}\mu\gamma}(\omega^{0})=-\frac{i}{16\pi^{2}}\frac{eg^{2}}{4M_{Z_{H}}^{2}}
(V_{Hl})_{ie}^*(V_{Hl})_{i\mu}(D+E+F)\\&& \quad\quad\ \
D=A(B_{0}^{a}\rightarrow\ B_{0}^{c},B_{1}^{a}\rightarrow\
B_{1}^{c})\\&& \quad\quad\ \  E=B(B_{0}^{b}\rightarrow\
B_{0}^{d},B_{1}^{b}\rightarrow\ B_{1}^{d})\\&& \quad\quad\ \
F=C(C_{\alpha\beta}^{1}\rightarrow\
C_{\alpha\beta}^{2},C_{\alpha}^{1}\rightarrow\
C_{\alpha}^{2},C_{0}^{1}\rightarrow\ C_{0}^{2}),\\&&
\Gamma^{\mu}_{\bar{e}\mu\gamma}(\omega^{\pm})=-\frac{i}{16\pi^{2}}\frac{eg^{2}}{2M_{W_{H}}^{2}}
(V_{Hl})_{ie}^*(V_{Hl})_{i\mu} [G +H -J]\\&& \quad\quad\ \ \
J=\{m_{Hi}^{2}m_{e}(p_{e}^{\mu}C_{0}^{4}+p_{\bar{\mu}}^{\mu}C_{0}^{4}+2C_{\mu}^{4})P_{L}
+m_{Hi}^{2}m_{\mu}(p_{e}^{\mu}C_{0}^{4}+p_{\bar{\mu}}^{\mu}C_{0}^{4}+2C_{\mu}^{4})P_{R}\\&&
\quad \quad\quad\quad\ \ \
-m_{Hi}^{2}[p_{\mu\alpha}(p_{e}^{\mu}+p_{\bar{\mu}}^{\mu}+2C_{\mu}^{4})+(p_{e}^{\mu}
+p_{\bar{\mu}}^{\mu})C_{\alpha}^{4}+2C_{{\mu}\alpha}^{4}]\gamma^{\alpha}P_{L}\\&&\quad
\quad\quad\quad\ \ \
-m_{e}m_{\mu}[p_{e\alpha}(p_{e}^{\mu}+p_{\bar{\mu}}^{\mu}+2C_{\mu}^{4})+(p_{e}^{\mu}
+p_{\bar{\mu}}^{\mu})C_{\alpha}^{4}+2C_{{\mu}\alpha}^{4}]\gamma^{\alpha}P_{L}\}~~~~~~~~~~~~~~~~~~~~~~~~~~~~~~~~~~~~~
\end{eqnarray*}
\begin{eqnarray*}
&&\quad \ \ \quad G=A(B_{0}^{a}\rightarrow\
B_{0}^{e},B_{1}^{a}\rightarrow\ B_{1}^{e})\\&&\quad \ \ \quad
H=B(B_{0}^{b}\rightarrow\ B_{0}^{f},B_{1}^{b}\rightarrow\
B_{1}^{f}),\\&&
\Gamma^{\mu}_{\bar{e}\mu\gamma}(A_{H})=-\frac{i}{16\pi^{2}}\frac{eg^{\prime2}}{50}
(V_{Hl})_{ie}^*(V_{Hl})_{i\mu}(K+L+M)
\\&&
\quad\quad \quad K=\frac{1}{p_{e}^{2}-m_{\mu}^{2}}[p_{e}^{2}
B_{1}^{a}+m_{\mu}\pslash_eB_{1}^{a} ]\gamma^{\mu}P_{L}\\
&&\quad\quad\ \quad
L=\frac{1}{p_{\bar{\mu}}^{2}-m_{e}^{2}}[p_{\bar{\mu}}^{2}
B_{1}^{b}-m_{e}\pslash_{\bar{\mu}}B_{1}^{b} ]\gamma^{\mu}P_{L}\\
&&\quad\quad\ \ \
M=[(\pslash_e+\pslash_{\bar{\mu}})C_{\alpha}^{1}\gamma^{\mu}\gamma^{\alpha}
-m_{Hi}^{2}C_{0}^{1}\gamma^{\mu}-C_{\alpha\beta}^{1}\gamma^{\alpha}\gamma^{\mu}\gamma^{\beta}
]P_{L},\\&&
\Gamma^{\mu}_{\bar{e}\mu\gamma}(Z_{H})=-\frac{i}{16\pi^{2}}\frac{eg^{2}}{2}
(V_{Hl})_{ie}^*(V_{Hl})_{i\mu}(N+O+P)\\&& \quad\quad\ \ \
N=K(B_{1}^{a}\rightarrow\ B_{1}^{c})\\&&\quad\quad\ \ \
O=L(B_{1}^{b}\rightarrow\ B_{1}^{d})\\&& \quad\quad\ \ \
P=M(C_{\alpha\beta}^{1}\rightarrow\
C_{\alpha\beta}^{2},C_{\alpha}^{1}\rightarrow\
C_{\alpha}^{2},C_{0}^{1}\rightarrow\ C_{0}^{2}),\\&&
\Gamma^{\mu}_{\bar{e}\mu\gamma}(W_{H}^{\pm})=-\frac{i}{16\pi^{2}}\frac{eg^{2}}{2}
(V_{Hl})_{ie}^*(V_{Hl})_{i\mu} (2Q+2R-T)\\&& \quad\quad\ \ \ \
T=[(p_{e}^{2}+2B_{0}^{f}+2m_{WH}^{2}C_{0}^{4})\gamma^{\mu}P_{L}
+(4C_{\alpha\mu}^{4}+2C_{\alpha}^{4}p_{b}^{\mu}
+2C_{\alpha}^{4}p_{\bar{\mu}}^{\mu})\gamma^{\alpha}P_{L}
+2\pslash_e(C_{\mu}^{4}\\&& \quad \quad \quad\quad\ \ \ \
+\pslash_ep^{b}_{{\bar\mu}}C_{0}^{4}+p_{\bar{\mu}}^{\mu}C_{0}^{4})P_{L}
+(2C_\alpha^{4}\pslash_e\gamma^{\alpha}
+C_\alpha^{4}\gamma^{\alpha}\pslash_e
+2C_\alpha^{4}\pslash_{\bar{\mu}}\gamma^{\alpha}+2\pslash_{\bar{\mu}}\pslash_eC_{0}^{4})\gamma^{\mu}P_{L}]
\\&&
\quad\quad\ \ \ \ Q=K(B_{1}^{a}\rightarrow\ B_{1}^{e})\\&&
\quad\quad\ \ \ \ R=L(B_{1}^{b}\rightarrow\ B_{1}^{f}),\\&&
\Gamma^{\mu}_{\bar{e}\mu\gamma}(W_{H}^{\pm}\omega^{\pm})=\frac{i}{16\pi^{2}}\frac{g^{2}e}{2}
(V_{Hl})_{ie}^*(V_{Hl})_{i\mu}\\&&\quad \quad \quad \quad \quad
\quad \quad
 \times
[m_{\mu}(\gamma^{\mu}\pslash_eC_{0}^{4}+\gamma^{\mu}\gamma^{\alpha}C_{\alpha}^{4})P_{R}
+m_{e}(\gamma^{\mu}\pslash_eC_{0}^{4}+\gamma^{\mu}\gamma^{\alpha}C_{\alpha}^{4})P_{L}
].\\&& \Gamma^{\mu}_{\bar{e}\mu
Z}(p_e,p_{\bar{\mu}})=\Gamma^{\mu}_{\bar{e}\mu
Z}(\eta)+\Gamma^{\mu}_{\bar{e}\mu Z}(\omega^{0})
+\Gamma^{\mu}_{\bar{e}\mu Z}(\omega^{\pm})+\Gamma^{\mu}_{\bar{e}\mu
Z}(A_{H})+\Gamma^{\mu}_{\bar{e}\mu Z}(Z_{H})
+\Gamma^{\mu}_{\bar{e}\mu Z}(W_{H}^{\pm})\\&&\quad \quad\quad
\quad\quad \ \ \quad+\Gamma^{\mu}_{\bar{e}\mu
Z}(W_{H}^{\pm}\omega^{\pm}),\\&& \Gamma^{\mu}_{\bar{e}\mu
Z}(\eta)=\frac{i}{16\pi^{2}}\frac{g}{\cos\theta_{W}}
(-\frac{1}{2}+\sin^{2}\theta_{W})\frac{g^{\prime2}}{100M_{A_{H}}^{2}}
(V_{Hl})_{ie}^*(V_{Hl})_{i\mu}(A'+B'+C')\\&& \quad \ \quad
A'=\{\frac{1}{p_{e}^{2}-m_{\mu}^{2}}[(-\frac{1}{2}+\sin^{2}\theta_{W})
(m_{Hi}^{2}m_{\mu}^{2}B_{0}^{a}+m_{Hi}^{2}p_{e}^{2}B_{1}^{a})\gamma^{\mu}P_{L}\\&&\quad
\quad  \quad \ \quad
+\sin^{2}\theta_{W}m_{e}m_{\mu}(m_{Hi}^{2}B_{0}^{a}+p_{e}^{2}B_{1}^{a})\gamma^{\mu}P_{R}
+(-\frac{1}{2}+\sin^{2}\theta_{W})(m_{e}m_{Hi}^{2}B_{0}^{a}\\&&\quad
\quad \ \quad \quad
+m_{e}m_{\mu}^{2}B_{1}^{a})\pslash_e\gamma^{\mu}P_{L}
+\sin^{2}\theta_{W}m_{Hi}^{2}m_{\mu}(B_{0}^{a}+B_{1}^{a})\pslash_e\gamma^{\mu}P_{R}]\}\\&&
 \quad \ \quad B'=\{\frac{1}{p_{\bar{\mu}}^{2}-m_{e}^{2}}[(-\frac{1}{2}+\sin^{2}\theta_{W})(m_{Hi}^{2}m_{e}^{2}B_{0}^{b}
+m_{Hi}^{2}p_{\bar{\mu}}^{2}B_{1}^{b})\gamma^{\mu}P_{L}\\&&\quad
\quad \quad \ \quad
+\sin^{2}\theta_{W}m_{e}m_{\mu}(m_{Hi}^{2}B_{0}^{b}+p_{\bar{\mu}}^{2}B_{1}^{b})\gamma^{\mu}P_{R}
-\sin^{2}\theta_{W}m_{e}m_{Hi}^{2}(B_{0}^{b}\\&&\quad \quad \quad \
\quad +B_{1}^{b})\gamma^{\mu}\pslash_{\bar{\mu}}P_{L}-(-\frac{1}{2}
+\sin^{2}\theta_{W})m_{\mu}(m_{Hi}^{2}B_{0}^{b}
+m_{e}^{2}B_{1}^{b})\pslash_{\bar{\mu}}\gamma^{\mu}P_{R}]\}\\&&~~~~~~~~~~~~~~~~~~~~~~~~~~~~~~~
\end{eqnarray*}
\begin{eqnarray*}
 &&\quad \ \quad \ \ C'=(-\frac{1}{2}+\sin^{2}\theta_{W})C,\\&&
\Gamma^{\mu}_{\bar{e}\mu
Z}(\omega^{0})=\frac{i}{16\pi^{2}}\frac{g}{\cos\theta_{W}}
\frac{g^{2}}{4M_{Z_{H}}^{2}}
(V_{Hl})_{ie}^*(V_{Hl})_{i\mu}{m_{Hi}^{2}}(D'+E'+F')\\&&
 \quad \ \ \quad D'=A'(B_{0}^{a}\rightarrow\
B_{0}^{c},B_{1}^{a}\rightarrow\ B_{1}^{c})\\&& \quad \ \ \quad
E'=B'(B_{0}^{b}\rightarrow\ B_{0}^{d},B_{1}^{b}\rightarrow\
B_{1}^{d})\\&& \quad \ \ \quad
F'=C'(C_{\alpha\beta}^{1}\rightarrow\
C_{\alpha\beta}^{2},C_{\alpha}^{1}\rightarrow\
C_{\alpha}^{2},C_{0}^{1}\rightarrow\ C_{0}^{2}),\\&&
\Gamma^{\mu}_{\bar{e}\mu
Z}(\omega^{\pm})=\frac{i}{16\pi^{2}}\frac{g}{\cos\theta_{W}}
\frac{g^{2}}{2M_{W_{H}}^{2}}(V_{Hl})_{ie}^*(V_{Hl})_{i\mu}
(G'+H'+I'+J')\\&&
  \quad \ \ \ \quad G'=A'(B_{0}^{a}\rightarrow\
B_{0}^{e},B_{1}^{a}\rightarrow\ B_{1}^{e})\\&& \quad \ \ \ \quad
H'=B'(B_{0}^{b}\rightarrow\ B_{0}^{f},B_{1}^{b}\rightarrow\
B_{1}^{f})\\&& \quad \ \ \ \ \quad
I'=(C'(C_{\alpha\beta}^{1}\rightarrow
C_{\alpha\beta}^{3},C_{\alpha}^{1}\rightarrow\
C_{\alpha}^{3},C_{0}^{1}\rightarrow\ C_{0}^{3})\\&&\quad \ \ \ \quad
J'=\cos^{2}\theta_WJ,\\&& \Gamma^{\mu}_{\bar{e}\mu
Z}(A_{H})=\frac{i}{16\pi^{2}}\frac{g}{\cos\theta_{W}}
\frac{g^{\prime2}}{50}(V_{Hl})_{ie}^*(V_{Hl})_{i\mu}(K'+L'+M')\\&&
\quad \ \ \ \quad
K'=\frac{1}{p_{e}^{2}-m_{\mu}^{2}}[(-\frac{1}{2}+\sin^{2}\theta_{W})p_{e}^{2}
B_{1}^{a}+\sin^{2}\theta_{W}m_{\mu}\pslash_eB_{1}^{a}
]\gamma^{\mu}P_{L}\\&& \quad \ \ \ \quad
L'=\frac{1}{p_{\bar{\mu}}^{2}-m_{e}^{2}}[(-\frac{1}{2}+\sin^{2}\theta_{W})p_{\bar{\mu}}^{2}
B_{1}^{b}-\sin^{2}\theta_{W}m_{e}\pslash_{\bar{\mu}}B_{1}^{b}
]\gamma^{\mu}P_{L}\\&& \quad \ \ \quad
M'=(-\frac{1}{2}+\sin^{2}\theta_{W})[(\pslash_e+\pslash_{\bar{\mu}})C_{\alpha}^{1}\gamma^{\mu}\gamma^{\alpha}
-m_{Hi}^{2}C_{0}^{1}\gamma^{\mu}-C_{\alpha\beta}^{1}\gamma^{\alpha}\gamma^{\mu}\gamma^{\beta}
]P_{L},\\&& \Gamma^{\mu}_{\bar{e}\mu
Z}(Z_{H})=\frac{i}{16\pi^{2}}\frac{g}{\cos\theta_{W}}
\frac{g^{2}}{2} (V_{Hl})_{ie}^*(V_{Hl})_{i\mu}(N'+O'+P')\\&& \quad \
\ \ \quad N'=K'(B_{1}^{a}\rightarrow\ B_{1}^{c})\\&&\quad \ \ \
\quad O'=L'(B_{1}^{b}\rightarrow\ B_{1}^{d})\\&& \quad \ \ \ \quad
P'=M'(C_{\alpha\beta}^{1}\rightarrow\
C_{\alpha\beta}^{2},C_{\alpha}^{1}\rightarrow\
C_{\alpha}^{2},C_{0}^{1}\rightarrow\ C_{0}^{2}),\\&&
\Gamma^{\mu}_{\bar{e}\mu Z}(W_{H}^{\pm})=
\frac{i}{16\pi^{2}}\frac{g}{\cos\theta_{W}}{g^{2}}(V_{Hl})_{ie}^*(V_{Hl})_{i\mu}(Q'+R'+S'+T')\\&&
\quad \ \ \ \quad Q'=K'(B_{1}^{a}\rightarrow\ B_{1}^{e})\\&&\quad \
\ \ \quad R'=L'(B_{1}^{b}\rightarrow\ B_{1}^{f})\\&&\quad \ \ \
\quad S'=-\frac{1}{2}M(C_{\alpha\beta}^{1}\rightarrow\
C_{\alpha\beta}^{3},C_{\alpha}^{1}\rightarrow\ C_{\alpha}^{3},
C_{0}^{1}\rightarrow\ C_{0}^{3})\\&& \quad \ \ \ \quad
 T'=\cos^{2}\theta_{W}T,\\&& \Gamma^{\mu}_{\bar{e}\mu
Z}(W_{H}^{\pm}\omega^{\pm})=\frac{i}{16\pi^{2}}2g^{3}\cos\theta_{W}
(V_{Hl})_{ie}^*(V_{Hl})_{i\mu}\\&&\quad \quad\quad \quad\quad \qquad
 \times [
m_{\mu}(\gamma^{\mu}\pslash_eC_{0}^{4}+\gamma^{\mu}\gamma^{\alpha}C_{\alpha}^{4})P_{R}
+m_{e}(\gamma^{\mu}\pslash_eC_{0}^{4}+\gamma^{\mu}\gamma^{\alpha}C_{\alpha}^{4})P_{L}
].~~~~~~~~~~~~~~~~~~~~~~~~~~~~~~~~~~~
\end{eqnarray*}
  For the two-point and three-point standard loop functions
$B_0,B_1,C_0,~C_{ij}$ in the above expressions are defined as
\begin{eqnarray*}
C_{ij}^{1}=C_{ij}^{1}(-p_{e},-p_{\bar{\mu}},m_{Hi},M_{A_{H}},m_{Hi}),
C_{ij}^{2}=C_{ij}^{2}(-p_{e},-p_{\bar{\mu}},m_{Hi},M_{Z_{H}},m_{Hi}),~~\\
C_{ij}^{3}=C_{ij}^{3}(-p_{e},-p_{\bar{\mu}},m_{Hi},M_{W_{H}},m_{Hi}),
C_{ij}^{4}=C_{ij}^{4}(p_{e},p_{\bar{\mu}},M_{W_{H}},m_{Hi},M_{W_{H}}),~~~~\\
B^{a}=B^{a}(-p_{e},m_{Hi},M_{A_{H}}),
B^{b}=B^{b}(p_{\bar{\mu}},M_{Hi},M_{A_{H}}),~~~~~~~~~~~~~~~~~~~~~~~~~~~~~~~~\\
B^{c}=B^{c}(-p_{e},m_{Hi},M_{Z_{H}}),
B^{d}=B^{d}(p_{\bar{\mu}},M_{Hi},M_{Z_{H}}),~~~~~~~~~~~~~~~~~~~~~~~~~~~~~~~~\\
B^{e}=B^{e}(-p_{e},m_{Hi},M_{W_{H}}),
B^{f}=B^{f}(p_{\bar{\mu}},M_{Hi},M_{W_{H}}).~~~~~~~~~~~~~~~~~~~~~~~~~~~~~~~\\
\end{eqnarray*}

\end{document}